\documentclass{article}
\begin{document}
\setlength{\textheight}{20 cm}
\title{Thermal equilibrium under the influence of gravitation}
\author{E. Fischer}
\date{Auf der Hoehe 82, 52223 Stolberg, Germany}

\maketitle

\begin{abstract}
Gas clouds under the influence of gravitation in thermodynamic equilibrium
cannot be isothermal due to the Dufour effect, the energy flux induced by
density gradients. In galaxy clusters this effect may be responsible for most
of the "cooling flows" instead of radiative cooling of the gas.
\end{abstract}

Recent observations of galaxy clusters with high spatial resolution have
shown that in most of them the intracluster plasma is far from being
isothermal, exhibiting strong decrease of the plasma temperature in the
vicinity of individual large galaxies. This effect is commonly attributed to
"cooling flows", that means, increasing radiative energy loss with increasing
density of the gas, which is gravitationally attracted by the galaxies. But
several discussions (Voigt et al. \cite{fab}, Markevitch et al. \cite{marke})
have shown that most of the temperature gradients would be washed out by
thermal conduction, if the thermal conductivity of the plasma were of the
magnitude given in the classical book by Spitzer \cite{spitzer}.

It has been proposed that magnetic fields may reduce the thermal
conductivity. The existence of magnetic fields in the order of $0.1 - 1
\rm{\mu G}$ has been confirmed by various observations (see e.g. Dolag et al.
\cite{dolag}). But to produce the required reduction, ordered magnetic fields
of a still higher magnitude would be necessary, as the heat flux reduction
works only perpendicular to the magnetic field lines. Narayan and Medvedev
\cite{naray} have shown that fields fluctuating on a large range of length
scales can produce only minor changes in the thermal conductivity.

In all these discussions it is implicitly assumed that in thermodynamic
equilibrium conduction leads to a uniform temperature, though it is well
known that density gradients may cause a flux of energy and thus induce
temperature gradients, an effect known from the textbooks as "Dufour effect"
(see e.g. Hirschfelder et al. \cite{hirsch}).

Speaking of equilibrium means that the net fluxes of all properties
determining the state of a system are locally balanced. Normally in a gas we
associate this with constant pressure, density and temperature. But in a
system, which is influenced by volume forces such as gravitation, the
equilibrium state may well exhibit gradients of the state variables. Mass
flow equilibrium in these systems is obtained by the balance between the
gravitational force and the counteracting pressure gradient, which is related
to the density gradient by some equation of state.

But if there is a density gradient, there must be also a temperature
gradient, balancing the Dufour effect, to obtain zero net energy flux. Below
we will derive the equilibrium conditions for a system, which is fully
relaxed with respect to flux of mass, momentum and energy, within the scope
of a simplified kinetic model.

For this purpose we determine the net flux of mass and energy through a
control area $\Delta F$ under the condition that the particles in the gas
have a Maxwellian distribution everywhere, but with number density and mean
energy varying perpendicular to the control area. The net flux is calculated
then under the assumption, that any particle crossing the control area
retains its velocity and direction, which it has obtained in the last
collision, one mean free path from the surface, and in addition is subjected
to some acceleration from the force field. In the energy balance the gain and
loss of kinetic energy due to the changing gravitational potential will be
omitted, as it cancels out from the balance exactly, when the net particle
flux through the control area is zero.

Denoting the direction perpendicular to the surface as z and the angle
between this and the direction of particle motion as $\theta$, the flux of
some property $G$ through the surface is
\begin {equation}
\label{jj}
\textstyle{j=\int \Delta F \cos \theta (j^+ - j^-) d \Omega,}
\end {equation}
where $j^+$ and $j^-$ are the normal components of the flux of $G$ carried by
particles moving towards the surface in positive and negative $z$ direction
under the angle $\theta$. The integration has to be taken over half the solid
angle $\pi /2\ge \theta\ge 0$. Denoting the amount of property $G$ carried by
particles moving with velocity $v$ by $g(v)$, the flux components of $g(v)$
are
\begin{equation}
\label{j+}
j^+=(g(v)-\frac {dg(v)}{dz}\lambda \cos \theta )(v \cos \theta + b \frac
{\lambda}{v})
\end{equation}
\begin{equation}
\label{j-}
j^-=(g(v)+\frac {dg(v)}{dz}\lambda \cos \theta )(v \cos \theta - b \frac
{\lambda}{v})
\end{equation}
$\lambda$ is the mean free path, which may depend also on the velocity. The
last term denotes the acceleration of the particles during their flight from
the last collision to the control surface. Integrating eq.(\ref{jj}) over
$d\Omega$ yields the equation
\begin{equation}
\label{j}
j= -\pi \Delta F \left ( \lambda v
\frac{dg(v)}{dz}  -\frac {2b\lambda}{v} g(v)
\right )
\end{equation}

To determine mass and energy transport by particles moving with velocity $v$,
$g$ has to be set to
\begin{equation}
\label{gM}
g_{\rm M}= n(z) m f(v,z) dv, \hspace{.6cm} g_{\rm E}=n(z) m \frac{v^2}{2}
f(v,z) dv
\end{equation}
In case of a neutral gas $n(z)$ and $m$ are the density and mass of atoms,
$f(v,z)$ is the distribution function
\begin{equation}
\label{f}
f(v) dv=\frac{4}{\sqrt{\pi}}\frac{v ^2}{\beta ^3}e^{-\frac{v ^2}{\beta ^2}}
dv
\end{equation}
$\beta (z)$ being the abbreviation $\beta (z)=\sqrt {2kT(z)/m}$. To calculate
the quantity $dg/dz$ we also need the derivative
\begin{equation}
\label{fz}
\frac{df}{dz} = \frac{d\beta}{dz}
\frac{4}{\sqrt{\pi}}\left(\frac{2v ^4}{\beta ^6}\!-\!\frac{3v ^2}{\beta ^4}\right
)e^{-\frac{v ^2}{\beta ^2}} = \frac{1}{\beta}\frac{d\beta}{dz}\!\left
(\frac{2v ^2}{\beta ^2}\!-3\!\right)f(v)
\end{equation}

While in a neutral gas the parameter $b$ in eq.(\ref{j}) is the normal
gravitational acceleration, in a plasma the action of the force is somewhat
indirect. Energy transport is mediated preferably by electrons. Thus in
eqs.(\ref{gM}) $n(z)$ and $m$ are the electron density and mass. But in this
case acceleration by the external force is acting also on the ions, which
transfer the force to the electrons by Coulomb interaction. Thus the
parameter $b$ in eqs.(\ref{j+}) and (\ref{j-}) is not the gravitational
acceleration of free moving electrons. But for the equilibrium condition of
zero transport the absolute magnitude of $b$ is not relevant, when we only
want to know, under which conditions the net flux of electrons and the flux
of energy carried by them are in balance.

An additional difference between neutral gas and plasma results from the fact
that the collision cross section between neutral atoms is nearly independent
of the collision energy, so that the mean free path $\lambda$ does not depend
on $v$. In plasmas, where Coulomb interaction is dominant, the mean free path
increases with $v^2$. Thus in the further calculations we set
$\lambda=\lambda _0 (v/v_0)^{\alpha}$ with $\alpha =0$ for neutral gas and
$\alpha =2$ for a fully ionised plasma.

Introducing eqs.(\ref{gM}) to (\ref{fz}) into eq.(\ref{j}), we get for the
mass and energy flux
\begin{equation}
\label{jM}
j_{\rm M}=2 \beta^{\alpha}
\left(x \beta \frac{dn}{dz} +
 x(2x^2-3) n\frac{d\beta }{dz}-\frac{2bn}{\beta }\frac{1}{x}\right)h(x)dx
\end{equation}
\begin{equation}
\label{jE}
j_{\rm E}=\beta^{\alpha +2}\!\!\left( x^3 \beta \frac{dn}{dz} +x^3 (2x^2\!-3)
n\frac{d\beta }{dz}\!-\!\frac{2bn}{\beta } x \right ) h(x)dx
\end{equation}
with $x=v/\beta$ and $h(x)=2\sqrt{\pi}m \Delta F
(\lambda_0/v_0^{\alpha})x^{2+\alpha} e^{-x^2}$.
\\Integrating these equations over all x, we obtain the conditions for zero mass and energy flux:
\begin{equation}
\label{FOM}
\beta\frac{dn}{dz} F_{3+\alpha} +n\frac{d
\beta}{dz} [2 F_{5+\alpha} -3 F_{3+\alpha} ]-\frac{2bn}{\beta} F_{1+\alpha} =0
\end{equation}
\begin{equation}
\label{FOE}
\beta\frac{dn}{dz} F_{5+\alpha} +n\frac{d
\beta}{dz} [2 F_{7+\alpha} -3 F_{5+\alpha} ]-\frac{2bn}{\beta} F_{3+\alpha} =0
\end{equation}

The abbreviation $F_k$ stands for the integral $F_k=\int_0^{\infty} x^k
e^{-x^2} dx$. The different arguments of $F_k$ in eqs.(\ref{FOM}) and
(\ref{FOE}) can be eliminated by the recurrence formula $F_{k+2}=(k+1)/2
\times F_k$, so that we
finally obtain the equilibrium conditions for mass and energy transport:
\begin{equation}
\label{FM}
\frac{2+\alpha}{2} \beta \frac{dn}{dz}+ \frac{(2+\alpha)(
1+\alpha )}{2} n \frac{d \beta}{dz}-\frac{2bn}{\beta}=0
\end{equation}
\begin{equation}
\label{FE}
\frac{4+\alpha}{2} \beta \frac{dn}{dz}+ \frac{(4+\alpha)( 3+\alpha )}{2} n
\frac{d \beta}{dz}-\frac{2bn}{\beta}=0
\end{equation}

It is immediately evident that this is a set of linear equations for the
gradients $d \beta /dz $ and $dn/dz$. If there are no external forces
$(b=0)$, there exists only the trivial solution $d\beta /dz=0$ and $dn/dz=0$.

In the presence of volume forces such as gravitation eqs.(\ref{FM}) and
(\ref{FE}) are inhomogeneous and allow a non-trivial solution, which
constitutes a fixed relation between temperature gradient and particle
density gradient or, because of quasi-neutrality, also between temperature
and mass density gradient. Eliminating $b$ leads to the relation
\begin{equation}
\beta \frac{dn}{dz}+(2\alpha+5) n \frac{d \beta }{dz}=0
\end{equation}
with the solution $ n \beta^{(2\alpha+5)}= \rm{const.}$ Replacing the number
density by the mass density $\rho$ and $\beta$ by $\sqrt{2kT/m}$ we finally
find the condition $\rho T^{5/2}= \rm{const.}$ for neutral gas and $\rho
T^{9/2}= \rm{const.}$ for a fully ionised plasma.

Though the relative temperature gradients are small compared to the
associated density gradients, they may well be important in the hot corona of
stars or in the intergalactic gas in galaxy clusters, where temperature
changes by a factor of three are reported (Voigt et al. \cite{fab}), while
the gas density may change by a few orders of magnitude in the vicinity of
large galaxies. It may even be that most of the observed temperature drop in
the "cooling flows" around these galaxies is not caused by radiative cooling
but by the "conductive cooling" associated with the density gradients.

It should be noted that due to the Dufour effect complete galaxy clusters
would have cooled off, if thermal conductivity were at the Spitzer value,
unless the clusters are embedded in a very hot but thin intergalactic plasma.
The existence of a diffuse high energy x-ray background (Boldt \cite{boldt})
is a strong hint to its existence. Matter expelled in the jets of quasars may
be the origin of this plasma.

\end{document}